\documentclass{aa}

\usepackage{graphicx}
\usepackage{txfonts}

\usepackage{amsmath}	% Advanced maths commands
\usepackage{amssymb}	% Extra maths symbols
\usepackage{mathtools}

\usepackage{color}
\usepackage{natbib,twoopt}
\usepackage[hyphenbreaks]{breakurl}
\usepackage[breaklinks]{hyperref}      %% to avoid \citeads line fills, add "draft" 
\bibpunct{(}{)}{;}{a}{}{,}             %% natbib format for A&A and ApJ
\definecolor{cobalt}{rgb}{0.06, 0.2, 0.65}
\hypersetup{
  colorlinks,
  citecolor=blue,
  linkcolor=blue,
  urlcolor=blue,
}
\makeatletter
  \newcommandtwoopt{\citeads}[3][][]{\href{http://adsabs.harvard.edu/abs/#3}%
    {\def\hyper@linkstart##1##2{}%
     \let\hyper@linkend\@empty\citealp[#1][#2]{#3}}}
  \newcommandtwoopt{\citepads}[3][][]{\href{http://adsabs.harvard.edu/abs/#3}%
    {\def\hyper@linkstart##1##2{}%
     \let\hyper@linkend\@empty\citep[#1][#2]{#3}}}
  \newcommandtwoopt{\citetads}[3][][]{\href{http://adsabs.harvard.edu/abs/#3}%
    {\def\hyper@linkstart##1##2{}%
     \let\hyper@linkend\@empty\citet[#1][#2]{#3}}}
  \newcommandtwoopt{\citeyearads}[3][][]%
    {\href{http://adsabs.harvard.edu/abs/#3}
    {\def\hyper@linkstart##1##2{}%
     \let\hyper@linkend\@empty\citeyear[#1][#2]{#3}}}
\makeatother

\newcommand{\Msun}{M$_{\odot}$}

\definecolor{smalt(darkpowderblue)}{rgb}{0.0, 0.2, 0.6}
\definecolor{forestgreen(traditional)}{rgb}{0.0, 0.5, 0.0}
\defcitealias{RVJ}{RVJ}

\defcitealias{MD17}{MD17}
\newcommand{\MDt}{\citetalias{MD17}}

\begin{document}

   %\title{Evidence for saturated and disrupted magnetic braking from samples of detached close binaries with low-mass main-sequence stars}

   \title{Evidence for saturated and disrupted magnetic braking from samples of detached close binaries with M and K dwarfs}

   %\subtitle{...........}

   \titlerunning{Evidence for saturated and disrupted magnetic braking}

   \author{Diogo Belloni\inst{1}
           \and
           Matthias R. Schreiber\inst{1,2}
           \and
           Maxwell Moe\inst{3}
           \and
           Kareem El-Badry\inst{4}
           \and
           Ken J. Shen\inst{5}
          }

    \authorrunning{D. Belloni et al.}

   \institute{Departamento de F\'isica, Universidad T\'ecnica Federico Santa Mar\'ia, Av. España 1680, Valpara\'iso, Chile\\
              \email{diogobellonizorzi@gmail.com}
              \and
              Millenium Nucleus for Planet Formation, Valpara{\'i}so, Chile
              \and
              Department of Physics \& Astronomy, University of Wyoming, Laramie, WY 82071
              \and
              Department of Astronomy, California Institute of Technology, 1200 E. California Blvd., Pasadena, CA 91125, USA
              \and
              Department of Astronomy and Theoretical Astrophysics Center, University of California, Berkeley, CA 94720, USA
             }

   \date{Received...; accepted ...}

% \abstract{}{}{}{}{} 
% 5 {} token are mandatory
 
  \abstract
   % context heading (optional), leave it empty if necessary  
   {
   Recent observations of close detached eclipsing M and K dwarf binaries have provided substantial support for magnetic saturation when stars rotate sufficiently fast, leading to a magnetic braking (MB) torque proportional to the spin of the star.
   }
   {
   We investigated here how strong MB torques need to be to reproduce the observationally-inferred relative numbers of white dwarf plus M dwarf post-common-envelope binaries under the assumption of magnetic saturation.
   }
   {
   We carried out binary population simulations with the BSE code adopting empirically-derived inter-correlated main-sequence binary distributions as initial binary populations and compared the simulation outcomes with observations.
   }
   {
   We found that the dearth of extreme mass ratio binaries in the inter-correlated initial distributions is key to reproduce the large fraction of post-common-envelope binaries hosting low-mass M dwarfs (${\sim0.1-0.2}$~\Msun). In addition, orbital angular momentum loss rates due to MB should be high for M dwarfs with radiative cores and orders of magnitude smaller for fully convective stars to explain the observed dramatic change of the fraction of short-period binaries at the fully convective boundary.
   }
   {
   We conclude that saturated but disrupted, that is, dropping drastically at the fully convective boundary, MB can explain the observations of both close main-sequence binaries containing M and K dwarfs and post-common-envelope binaries. Whether a similar prescription can explain the spin down rates of single stars and of binaries containing more massive stars needs to be tested.
   }


   \maketitle
%
%-------------------------------------------------------------------

%=============================
%=============================
%        BODY
%=============================
%=============================

%%%%%%%%%%%%%%%%
% NEW SECTION
%%%%%%%%%%%%%%%%
\section{Introduction}
\label{introduction}

Understanding how a magnetized wind extracts angular momentum from a star, so called magnetic braking, is a key ingredient to understand the evolution of close binaries as important as cataclysmic variables (CVs), low-mass X-ray binaries, ultra-compact X-ray binaries, or double white dwarfs.   
Despite this importance, the strength and main dependencies of magnetic braking, in particular on the mass and rotation period of the star, remain puzzling.

In early studies magnetic braking was calibrated using the spin down rates of solar-type stars \citep{skumanich72-1} but recently it has become clear that, in particular for lower mass main-sequence stars, the situation is more complicated \citep[e.g.][]{Barnes_2003,newtonetal16-1} which most likely hints towards different and mass dependent magnetic braking laws.
One frequently discussed attempt to describe magnetic braking is based on the observation that chromospheric activity, coronal X-ray emission, flare activity, and magnetic field strengths in low-mass main-sequence stars are correlated and increase with rotation up to a mass dependent critical rotation rate above which the relation between activity and rotation saturates.
The assumption that these observables also relate with magnetic braking led to postulating saturated magnetic braking prescriptions in which the dependence of the magnetic braking torque on the spin period becomes shallower above a given rotation rate \citep[e.g.][]{Chaboyer_1995,Sills_2003,APS}.

In binaries with orbital periods shorter than $\sim5-10$~d \citep[e.g.][]{flemmingetal19-1}, the spin period is synchronised with the orbital period and magnetic braking therefore leads to orbital angular momentum loss.
Changes in the orbital period, or distributions of representative samples of close binary stars, which are in principle easier to measure than rotation rates of single stars, can therefore be used to constrain the dependencies and strength of magnetic braking.

In semi-detached binary stars with a white dwarf accreting from a main-sequence star companion, the previously mentioned CVs, magnetic braking is often assumed to be absent (or very weak) in case the donor star is fully convective but very efficient in case it still has a radiative core.
The prescription used for binary population models of CVs, called disrupted magnetic braking, assumes efficient magnetic braking from \citet{RVJ}, which is based on the \citet[][]{skumanich72-1} prescription, combined with the assumption that magnetic braking becomes inefficient as soon as the main-sequence star becomes fully convective.  
These assumptions reasonably well explain the increased radii of the main-sequence stars at long orbital periods ($\sim3-5$ hr), a dearth of systems in the orbital period range of $\approx2-3$\,hr, and most of the mass transfer rates derived from observations below the period gap \citep[e.g.][]{Knigge_2011_OK,McAllister_2019,palaetal_2017,Pala_2022}.

However, CVs are far from ideal systems to constrain magnetic braking because of two main reasons.
First, mass transfer may be increased by so-called consequential angular momentum loss \citep{Schreiber_2016} which is not well understood and contaminates the relation between mass transfer rates and angular momentum loss due to magnetic braking.
Second, mass transfer rates are difficult to measure and the few measurements that are available show quite a large scatter at a given orbital period.
It might therefore not be too surprising that the current model for CV evolution fails to explain some important observables \citep[e.g.][]{Belloni_2020a,Pala_2020,FuentesMorales_2021,Pala_2022,BelloniSchreiber2023hxgabook}.

Instead of using single stars or CVs we here combine observational constraints from the two cleanest and most suitable types of system towards a better understanding of magnetic braking. 
The first has been provided by \citet{schreiberetal10-1}.
They observed a large number of detached binaries consisting of a white dwarf with an M dwarf companion and find a strong dependence of the relative number of short orbital period systems, which are post-common-envelope binaries (PCEBs), on the mass of the main-sequence star.
This measurement is very clean because the orbital period evolution of these systems is not affected by mass transfer, the masses of the stellar components can relatively easily be estimated, and we know that the orbital period distribution of the PCEBs peaks at a few hours and that there are very few systems with orbital periods exceeding one day \citep{nebot-gomez-moranetal11-1}.
Using these systems to constrain angular momentum loss through magnetic braking has been suggested more than a decade ago \citep[e.g.][]{Politano_2006,Zorotovic_2010} but no dedicated simulations have ever been performed. 

The second clean observational constraint that we take into account here comes from eclipsing close main-sequence binaries \citep{ElBadry_2022}. 
The observed orbital period distributions of these systems provide evidence for a magnetic braking torque that has a shallower dependence on the star spin than assumed by \citet{RVJ} and can be reasonably well understood assuming saturated magnetic braking laws.
While the eclipsing main-sequence binary star sample is very useful to constrain the dependence of magnetic braking on the spin period, it is less sensitive to its strength or possible dependencies on the stellar mass. 
In this work we investigate if a prescription of (saturated) magnetic braking exists that can explain these two critical observational constraints from close detached binary stars using binary population synthesis.

%%%%%%%%%%%%%%%%%%%%%%%%%%%%%%%%
%%%%%%%%%%%%%%%%%%%%%%%%%%%%%%%%
% NEW SECTION
%%%%%%%%%%%%%%%%%%%%%%%%%%%%%%%%
%%%%%%%%%%%%%%%%%%%%%%%%%%%%%%%%
\section{Initial binary populations}
\label{IBP}

In previous population models of white dwarf plus main-sequence star binaries (PCEBs and CVs), the initial binary population (IBP) has been based on uncorrelated distributions of initial parameters for the main-sequence binaries \citep[e.g.][]{Kool_1992,Kool_1993,Willems_2004,Politano_2006,Davis_2010,Toonen_2013,Zorotovic_2014,Camacho_2014,Schreiber_2016,Cojocaru_2017,Belloni_2018b,Belloni_2020a}.
However, assumptions for uncorrelated distributions are not consistent with observational surveys performed in the past decades.
In addition, the measurement we aim to compare with predictions in this work is the fraction of PCEBs among white dwarf plus M dwarf binaries identified by the Sloan Digital Sky Survey \citep{schreiberetal10-1}, that is, the observable we want to explain is the number of close binaries that evolved through common-envelope evolution among all spectroscopically identified white dwarf plus M dwarf binaries, which includes wide (but still unresolved) systems. 
Therefore, the number of wide systems, that is, systems in which the two stars are so far apart that they never interacted, is as important as the number of PCEBs which implies that we need to consider a large range of initial orbital periods.

\citet[][hereafter MD17]{MD17} and \citet{Offner_2023} investigated dozens of surveys related to main-sequence binaries and, after combining the samples from such surveys and correcting for their respective selection effects, concluded that the distributions of periods, masses, and mass ratios are not independent at a statistically significant level and fitted joint probability density functions ${f(M_1,q,P_{\rm orb},e) \neq f(M_1)f(q)f(P_{\rm orb})f(e)}$ to the corrected distributions, where $M_1$ is the primary mass, $q=M_2/M_1$ is the mass ratio and $M_2$ is the secondary mass, $P_{\rm orb}$ is the orbital period and $e$ is the eccentricity. 
These fitted correlated distributions are the most realistic ones currently available and should be incorporated into binary population models.
In addition, we expect them to be crucial if a large range of initial parameters play an important role as in the case of the present investigations.
We incorporated the distributions published by \MDt~and compared the resulting predictions with those of the standard ``flat''uncorrelated distributions and the observations.

%----------------------
% NEW SUBSECTION
%----------------------
\subsection{Standard ``flat'' uncorrelated distributions}
\label{flat}

The IBPs commonly adopted in population synthesis related to PCEBs and CVs are very simplistic, as the focus is often on testing parameters of binary evolution, rather than inspecting how the assumed initial distributions might affect the results.
Apart from the initial mass function, distributions for the mass ratio, separation and eccentricity are usually assumed to be flat, being the separation flat in log-scale and the eccentricity flat in squared-scale (i.e. thermal eccentricity distribution).
As we intend to test the impact of correlated distributions derived by \MDt~(Sect.~\ref{MD17}), we also perform binary population synthesis with these commonly adopted flat initial distributions.

Our ``flat'' IBP corresponds to ${\approx6.15\times10^5}$ zero-age main-sequence binaries characterized by:
(i) $M_1$ is obtained from the canonical \citet{Kroupa_2001} initial mass function, in the range $[1,8]$ M$_\odot$, as most white dwarf progenitors belong to this mass range;  
(ii) $M_2$ is obtained assuming a uniform mass ratio distribution, where $M_2 \leq M_1$, and requesting that $M_2\geq0.069$, in order to avoid sub-stellar secondaries; 
(iii) the semi-major axis $(a)$ follows a log-uniform distribution in the range between $1.1$ times the sum of the radii of the two stars and $10^{5.5}$~R$_\odot$; and  
(iv) the eccentricity is uniform in $e^2$, which means that $e$ follows a thermal distribution in the range $[0,1]$.

%----------------------
% NEW SUBSECTION
%----------------------
\subsection{Inter-correlated distributions of \citet{MD17}}
\label{MD17}

In our second IBP, we also selected ${\approx6.15\times10^5}$ zero-age main-sequence binaries and picked $M_1$ from the canonical \citet{Kroupa_2001} initial mass function in the range between $1$ and $8$~\Msun. 
However, we adopted the correlated distributions derived by \MDt, in which the $P_{\rm orb}$ distribution depend critically on $M_1$ and the binary fraction and $e$ and $q$ distributions depend on both $P_{\rm orb}$ and $M_1$.

We incorporate the frequency $f_{{\rm log}P_{\rm orb};q>0.1}(M_1,P_{\rm orb})$ of companions with mass ratio ${q>0.1}$ per decade of $P_{\rm orb}$ from \MDt, but made three slight modifications for the current study.
First, we convolve their analytic relations for $f_{{\rm log}P_{\rm orb};q>0.1}$ with a Gaussian 2D kernel of width ${\delta\log({P_{\rm orb}/{\rm d})}=0.4}$ and ${\delta\log({M_1}/{\rm M}_\odot)=0.1}$ to ensure smooth transitions across the 
parameter space.
Second, we interpolate between their value of $f_{{\rm log}P_{\rm orb};q>0.1}$ at ${P_{\rm orb}=2}$~d and zero at the minimum possible orbital period, which corresponds to Roche-lobe filling on the zero-age main sequence.
Finally, we extend their distributions toward slightly longer periods ${\log(P_{\rm orb}/{\rm d})=9}$.

For ${M_1=1}$~\Msun~primaries, the binaries follow a log-normal period distribution with peak of ${f_{{\rm log}P_{\rm orb};q>0.1}=0.12}$ at ${\log(P_{\rm orb}/{\rm d})=5}$ (${\approx50}$~au) and a value of ${f_{{\rm log}P_{\rm orb};q>0.1}=0.03}$ at ${\log(P_{\rm orb}/{\rm d})=1}$.
By integrating ${f_{{\rm log}P_{\rm orb};q>0.1}}$ across all orbital periods, the mean multiplicity frequency of companions per ${M_1=1}$~\Msun~primary is ${f_{\rm mult}=0.58}$.
This orbital period distribution and multiplicity frequency are consistent with previous parameterizations of solar-type binaries \citep{DM91,Raghavan_2010}.

Meanwhile, the frequency of companions to more massive primaries is larger, especially at short orbital periods.
For ${M_1=8}$~\Msun, we find a peak of ${f_{{\rm log}P_{\rm orb};q>0.1}=0.28}$  at ${\log(P_{\rm orb}/{\rm d})=4}$ (${\approx10}$~au), a value of ${f_{{\rm log}P_{\rm orb};q>0.1}=0.16}$ at ${\log(P_{\rm orb}/{\rm d})=1}$, and a multiplicity frequency of  ${f_{\rm mult}=1.6}$.
These statistics are consistent with the observed properties of B-type main-sequence binaries \citep{Abt_1990,Rizzuto_2013}.
A significant fraction of PCEBs evolve from main-sequence binaries with intermediate orbital periods ${\log{\rm (P_{\rm orb}/d)}=2-4}$.  
Across this period interval, the frequency of companions increases by nearly a factor of four between ${M_1=1}$~\Msun~and ${M_1=8}$~\Msun~primaries.

\begin{figure}
   \begin{center}
    \includegraphics[width=0.99\linewidth]{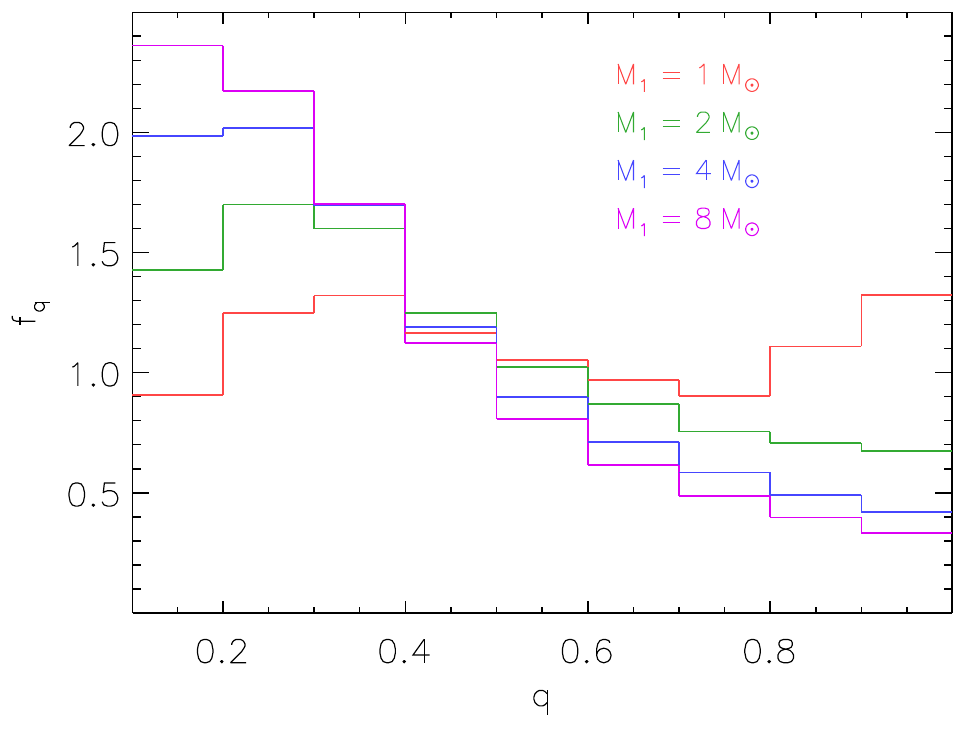}
    \end{center}
  \caption{Mass-ratio distribution of zero-age main-sequence binaries with orbital period ${P_{\rm orb}=10^2-10^4}$~d, separated according to the primary mass $M_1$.
  }
  \label{Fig02}
\end{figure}

The power-law slope $\alpha$ of the eccentricity $e$ distribution ${f_e \propto e^{\alpha}}$ also depends on the primary mass and orbital period.
We adopt the parameterization for ${\alpha(M_1,P_{\rm orb}}$) as given in \MDt.
We also limit the distribution to the maximum eccentricity $e_{\rm max}$ without Roche-lobe filling at periastron.
Specifically, we adopt the power-law slope $\alpha$ across the interval ${0<e<0.7e_{\rm max}}$, and then model $f_e$ as a linearly declining function until it reaches zero at ${e=e_{\rm max}}$.

We adopt the three-component mass-ratio distribution $f_q$ as described in \MDt: a power-law slope across small mass ratios ${q=0.1-0.3}$, a power-law slope across large mass ratios ${q=0.3-1.0}$, and an excess fraction of twins.  
All three components of the mass-ratio distribution depend on $M_1$ and $P_{\rm orb}$.
When modelling PCEBs, we are particularly concerned with the mass-ratio distribution across intermediate orbital periods ${P_{\rm orb}=10^2-10^4}$~d, which we display in Fig.~\ref{Fig02}.
In the current study, we model the twin component as a function that quickly rises from ${q=0.85}$ and peaks at ${q=1.0}$ \citep{Tokovinin_2000,Halbwachs_2003,Moe_2013}.
The mass-ratio distribution of solar-type binaries with intermediate orbital periods is close to flat, possibly with a small excess at ${q=0.3}$ and ${q>0.85}$ \citep{DM91,Halbwachs_2003,Raghavan_2010}.
Meanwhile, the mass-ratio distribution of A-type and B-type main-sequence binaries with intermediate orbital periods is considerably skewed towards ${q=0.3}$ with a flattening across ${q=0.1-0.3}$  \citep{Rizzuto_2013,Gullikson_2016,Murphy_2018}.
Combining all the statistics, there are ${\approx8}$ times more companions with $q=0.2$ and ${\log{(P_{\rm orb}/{\rm d})}=2-3}$ around ${M_1=8}$~\Msun~primaries compared to their counterparts around ${M_1=1}$~\Msun~primaries.

We incorporate the frequency $f_{{\rm log}P_{\rm orb};q>0.1}(M_1,P_{\rm orb})$ of companions with mass ratio ${q>0.1}$ per decade of $P_{\rm orb}$ as explained above.
For ${q<0.1}$, the statistics are unfortunately poorly constrained from observations.
At separations ${\lesssim1}$~au, there is a brown dwarf desert, that is, an extreme deficit of ${q<0.1}$ companions to both solar-type primaries  \citep{Grether_2006,Ma_2014} and A/early-F primaries \citep{Murphy_2018}.
Across separations ${\sim10-100}$~au, the companion mass distribution is relatively uniform across ${q=0.01-0.2}$, that is, there are about as many ${q=0.01-0.1}$ companions as there are ${q=0.1-0.2}$ companions \citep{Wagner_2019,Nielsen_2019}.
There are no constraints on ${q<0.1}$ companions to B stars except for beyond separations ${>100}$~au.
Finally, there are no good constraints on ${q<0.1}$ companions of any spectral type across separations ${\sim1-10}$~au, which is precisely the range from which most PCEBs should originate.

Binary formation models of disk fragmentation, accretion, and migration can reproduce many of the observed features of the initial binary population, including the brown dwarf desert at separations ${<1}$~au and the sizeable population of brown dwarf companions beyond ${>10}$~au \citep{Tokovinin_2020}.
However, the transition between these two regimes is substantially model dependent.
Given the incompleteness and selection effects of the observed samples and the systematic errors in the physical modelling, the occurrence rate of brown dwarf companions to B--G primaries across separations of ${\sim1-10}$~au is uncertain by a factor of ${\sim3}$.

We assume here an absence of companions with ${q<0.1}$ \citep[e.g.][]{Duchene_2023}, but bear in mind that some systems must have come from extreme mass ratio binaries.
In particular, there are several observed PCEBs hosting brown dwarfs \citep{Zorotovic_2022}, which could only be explained if they originated from zero-age main-sequence binaries with ${q<0.1}$.
Similarly, PCEBs with massive WDs (${\gtrsim1}$~\Msun) having M dwarf companions should come from zero-age main-sequence binaries with ${q<0.1}$ as the WD progenitors in these cases are initially more massive than ${\sim5}$~\Msun.
Despite that, the contribution of systems with ${q<0.1}$ to the total population of PCEBs is most likely negligible as the fraction of PCEBs with either high-mass WDs or brown dwarfs makes up only a few per cent \citep{Zorotovic_2010,nebot-gomez-moranetal11-1,Brown_2023}.

When compared with the ``flat'' uncorrelated IBP described in Sect.~\ref{flat}, the inter-correlated distributions of \citet{MD17} contain two improvements, which are crucial for the binary star populations we investigate in this paper.
The first one is the dearth of binaries with ${q<0.1}$ we have just described above.
The other one is the dependence of the binary properties on the primary mass.
In particular, with increasing primary mass, the binary fraction increases across separations of ${1-10}$~au and their mass-ratio distribution becomes skewed toward ${q\sim0.3}$, with the aforementioned turnover below ${q<0.1}$.
This implies that the fraction of binaries that evolve via common-envelope evolution with M-dwarf companions is substantially higher than in the other IBP.
The remaining features of the IBP from \citet{MD17} have little impact on the binary populations considered in this work.

%%%%%%%%%%%%%%%%%%%%%%%%%%%%%%%%
%%%%%%%%%%%%%%%%%%%%%%%%%%%%%%%%
% NEW SECTION
%%%%%%%%%%%%%%%%%%%%%%%%%%%%%%%%
%%%%%%%%%%%%%%%%%%%%%%%%%%%%%%%%
\section{Binary population models}
\label{BSE}

For both IBPs (Sect.~\ref{IBP}) we carried out binary population synthesis using the BSE code assuming solar metallicity (i.e. ${Z=0.02}$) and a constant star formation rate \citep[e.g.][]{Weidner_2004,Kroupa_2013,Recchi_2015,Schulz_2015} over the age of the Galactic disc \citep[${\approx10}$~Gyr,][]{Kilic_2017}.
For the common-envelope evolution, we adopted an efficiency of $0.25$, that is, we assumed that $25$\% of the change in orbital energy during the spiral-in is used to unbind the common envelope, with no contributions from other energy sources, which is consistent with the increasing evidence that PCEB progenitors experience strong orbital shrinkage during common-envelope evolution
\citep[e.g.][]{Zorotovic_2010,Toonen_2013,Camacho_2014,Cojocaru_2017,Belloni_2019,hernandezetal22-1, Zorotovic_2022,schrebak+fuller23-1}.
The binding energy parameter was calculated according to fitting scheme provided by \citet[][their Appendix A]{Claeys_2014}, which is based on the detailed numerical stellar evolution calculations by \citet{Dewi_2000} and takes into account the structure and the evolutionary stage of the red giant donor.
All other stellar and binary evolution parameters not clearly mentioned are set as the standard in BSE \citep[e.g.][]{Hurley_2002,Belloni_2018b,Banerjee_2020}.
Finally, in this work, we only considered PCEBs that are pre-CVs, that is, we do not consider detached CVs crossing the period gap.

After a PCEB is formed, it evolves towards shorter periods through orbital angular momentum loss.
In addition to magnetic braking, we also included emission of gravitational waves as mechanism to remove orbital angular momentum as described in \citet[][section 2.4, equation 48]{Hurley_2002}. 
Regarding magnetic braking, we adopted the following prescription with magnetic saturation, which was first proposed by \citet{Chaboyer_1995},

\begin{equation}
\dot{J}_{\rm MB,SAT} \, = \,
-\beta
\left(
  \frac{R_2}{{\rm R}_{\odot}} \,
  \frac{{\rm M}_{\odot}}{M_2}
\right)^{1/2}
\left\{
\begin{array}{ll}
\Omega_2^3,                     & {\rm if} \; \Omega_2 \leq \Omega_{\rm crit}, \\
\Omega_2\,\Omega_{\rm crit}^2,  & {\rm if} \; \Omega_2 > \Omega_{\rm crit},
\end{array}
\right.
\label{MBsat}
\end{equation}

\noindent
where ${\beta=2.7\times10^{47}}$~erg~s$^{-1}$ \citep{APS}, and $M_2$, $R_2$, and $\Omega_2$ are the mass, radius, and spin frequency (in s$^{-1}$) of the main-sequence star, respectively.
$\Omega_{\rm crit}$ is the threshold angular velocity beyond which saturation occurs and is assumed to be
\citep{ElBadry_2022}

\begin{equation}
\Omega_{\rm crit} \ = \ 
10\,\Omega_\odot
\left(
  \frac{\tau_{\odot}}{\tau_{\rm 2}}\,
\right),
\label{SCRIT}
\end{equation}

\noindent
where ${\Omega_\odot=3\times10^{-6}~{\rm s}^{-1}}$ and $\tau_{\rm 2}$ is the convective turnover time-scale of the main-sequence star given by \citep{Wright_2011}

\begin{equation}
\log_{10}
\left(
 \frac{\tau_{\rm 2}}{\rm d}
\right)
=
1.16
-1.49
\log_{10}
\left( 
  \frac{M_2}{{\rm M}_\odot}
\right)
-0.54
\log_{10}^2
\left( 
  \frac{M_2}{{\rm M}_\odot}
\right).
\label{TAU}
\end{equation}

According to Eqs.~\ref{TAU} and \ref{SCRIT}, the lower the mass of the main-sequence star, the longer the convective turnover time-scale and the longer the critical spin period below which magnetic braking is saturated.
In particular, the saturation spin period is as long as ${\sim21.6}$~d, for a $0.1$~\Msun~star, and as short as ${\sim2.85}$~d, for a $0.9$~\Msun~star.

To test different strength of magnetic braking and different levels of disrupted magnetic braking, we introduced two multiplicative factors. First, we add a factor $K$ with which we can scale the strength of magnetic braking. Second, for fully convective stars, that is, those less massive than ${\sim0.35}$~\Msun, we added an additional parameter $\eta$ to the expression such that magnetic braking is reduced by a factor of $\eta$ for these stars.

\begin{equation}
\dot{J}_{\rm MB} \, = \,
\left\{
\begin{array}{ll}
K \, \dot{J}_{\rm SAT} \, ,         & {\rm if} \; M_2   >  0.35\,{\rm M}_\odot, \\
 & \\
\left(K \, \dot{J}_{\rm SAT}\right)/\eta \, , & {\rm if} \; M_2 \leq 0.35\,{\rm M}_\odot ~ ({\rm fully~convective}).
\end{array}
\right.
\label{MBrecipe}
\end{equation}

We here focus on this prescription and test whether it can also explain the fraction of PCEBs amongst the entire population of white dwarf plus M dwarf binaries, and for which combination of $K$ and $\eta$.
Naturally, Eq.~\ref{MBrecipe} reduces to Eq.~\ref{MBsat} when ${K=1}$ and ${\eta=1}$.
In addition, magnetic braking becomes entirely disrupted for fully convective M dwarfs when ${\eta\rightarrow\infty}$ is assumed.

%%%%%%%%%%%%%%%%%%%%%%%%%%%%%%%%
%%%%%%%%%%%%%%%%%%%%%%%%%%%%%%%%
% NEW SECTION
%%%%%%%%%%%%%%%%%%%%%%%%%%%%%%%%
%%%%%%%%%%%%%%%%%%%%%%%%%%%%%%%%
\section{Results}
\label{Results}

\begin{figure*}
  \begin{center}
    \includegraphics[width=0.99\linewidth]{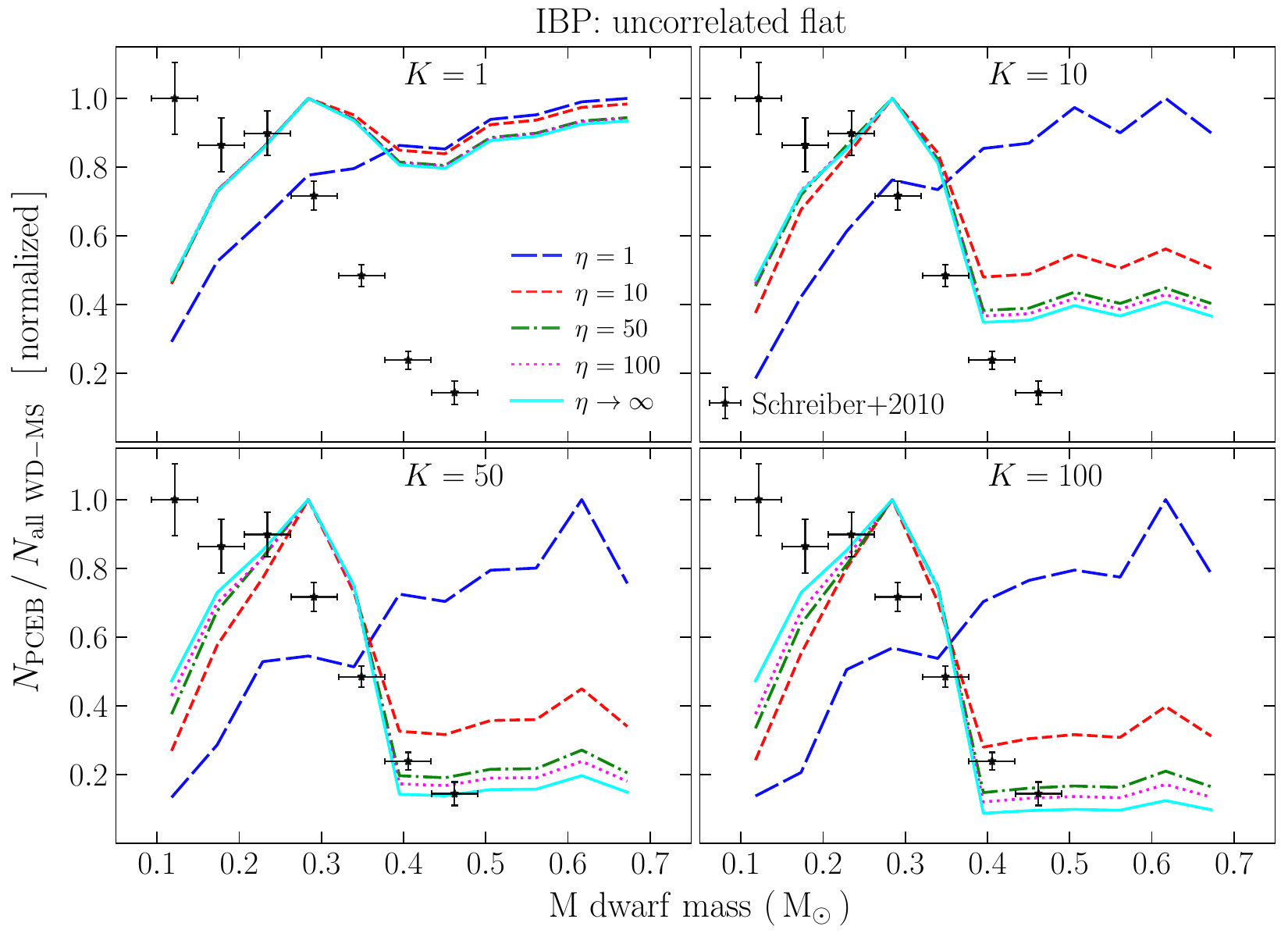}
  \end{center}
  \caption{Comparison between the observed fractions of PCEBs amongst white dwarf plus M dwarf binaries across the M dwarf mass \citep{schreiberetal10-1} and the predicted with Eq.~\ref{MBrecipe} for several combinations of model parameters. Both predicted and observed fractions were normalized at their highest values. Each panel corresponds to a different choice for $K$, while the line colour and type indicate the assumed value of $\eta$. We can see that any combination of the parameters such that ${K\gtrsim50}$ and ${\eta\gtrsim50}$ is able to reasonably well explain the high fraction of systems for M dwarf masses ${\lesssim0.3}$~\Msun~as well as the huge reduction of systems at $\sim0.5$~\Msun.}
  \label{FigPCEBflat}
\end{figure*}

\begin{figure*}
  \begin{center}
    \includegraphics[width=0.99\linewidth]{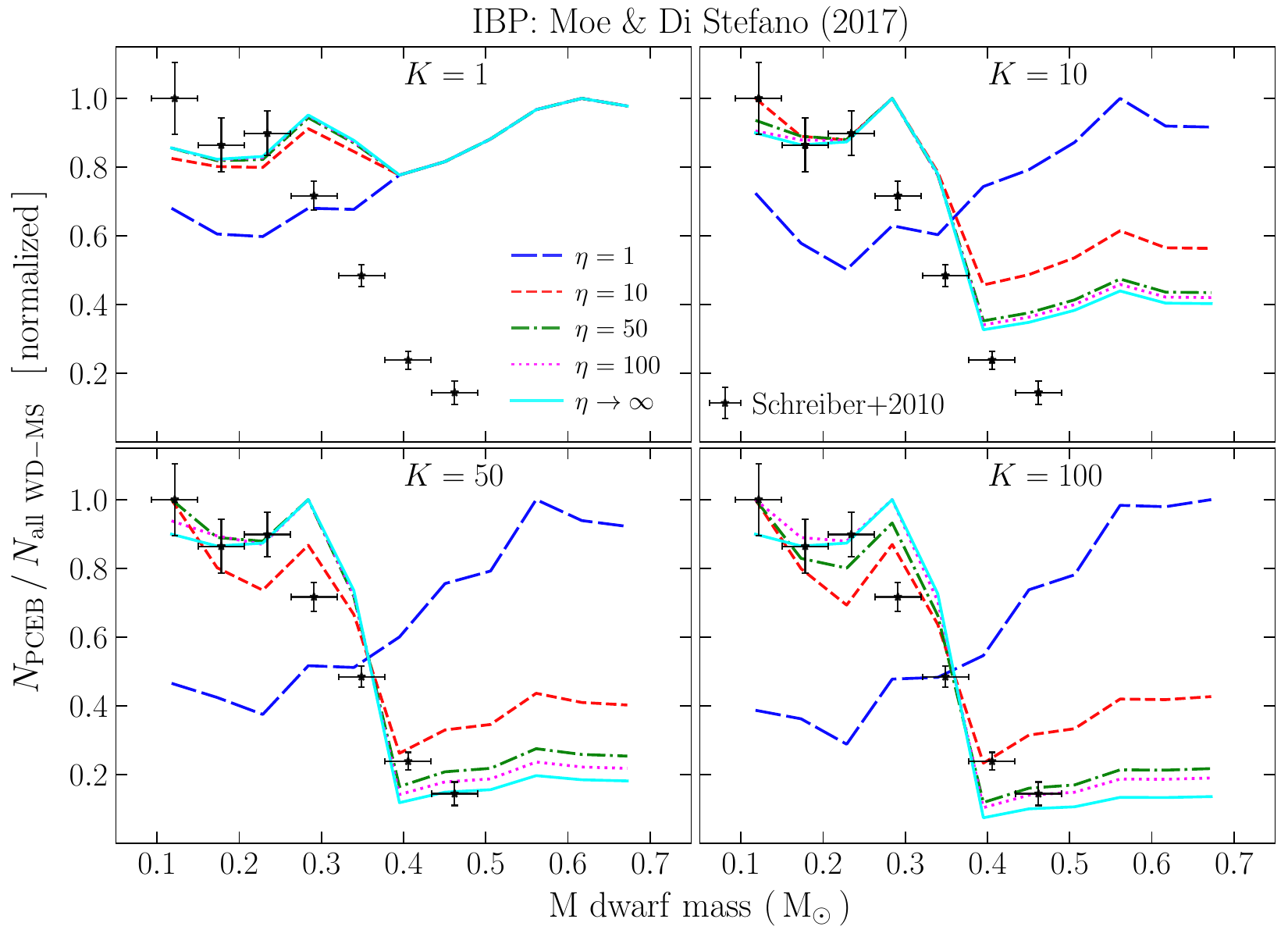}
  \end{center}
  \caption{Similar to \ref{FigPCEBflat}, but assuming the inter-correlated IBP derived by \MDt. Analogous to what we have found while adopting an IBP draw from flat distributions, any combination of the parameters such that ${K\gtrsim50}$ and ${\eta\gtrsim50}$ is able to reasonably well explain the huge reduction of systems at ${\sim0.5}$~\Msun. However, with a more realistic IBP, we can also reproduce the high fraction of systems for M dwarf masses ${\lesssim0.3}$~\Msun.}
  \label{FigPCEBmd17}
\end{figure*}

\begin{figure*}
  \begin{center}
    \includegraphics[width=0.98\linewidth]{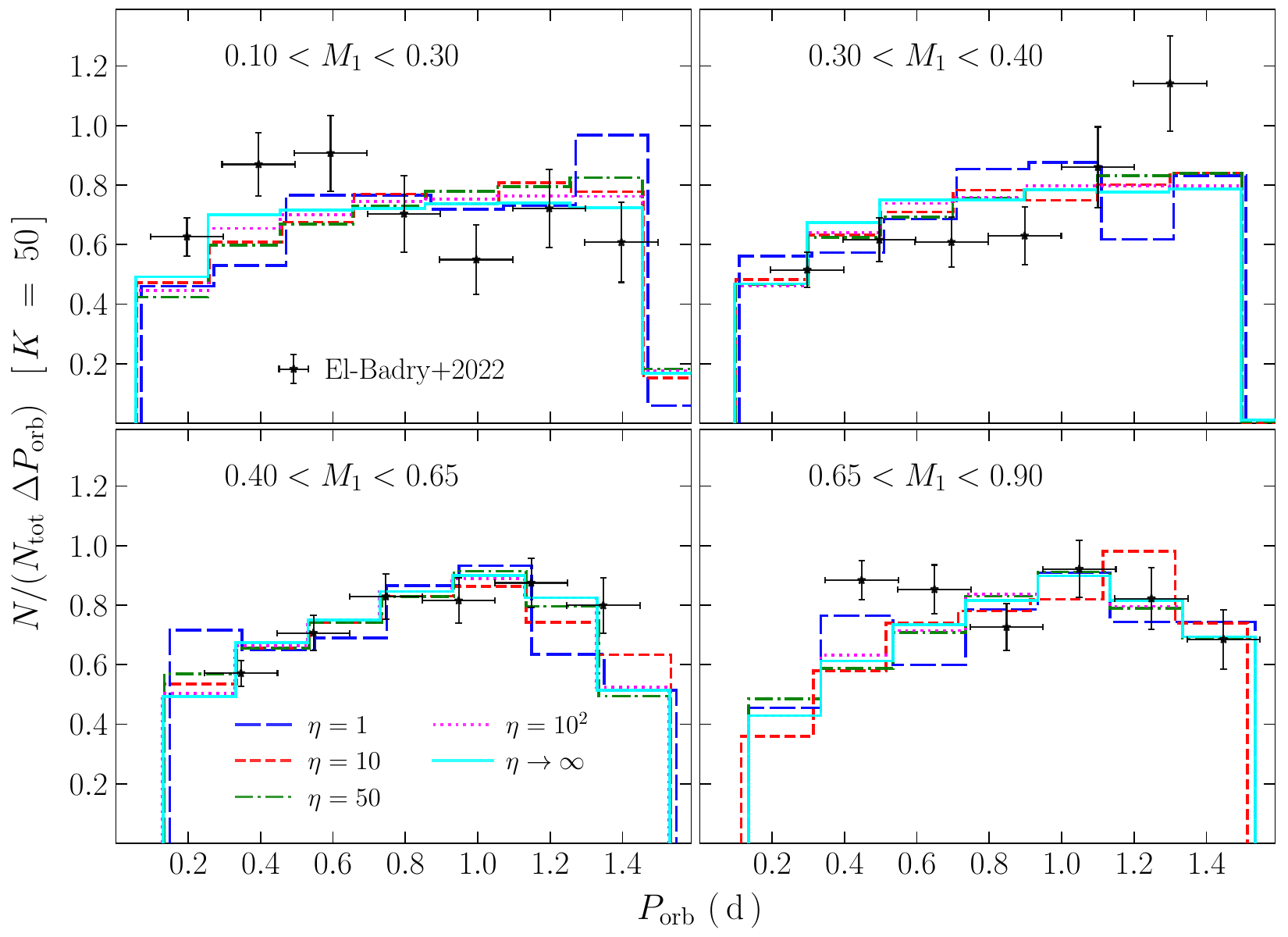}
  \end{center}
  \caption{Comparison between the observed orbital period distributions for different mass bins \citep{ElBadry_2022} and the predicted assuming ${K=50}$ and several values of $\eta$. The line types and colours indicate the assumed value of $\eta$, while each panel corresponds to a different primary mass bin, being ${0.10-0.30}$~\Msun~(top left panel), ${0.30-0.40}$~\Msun~(top right panel), ${0.40-0.65}$~\Msun~(bottom left panel), and ${0.65-0.90}$~\Msun~(bottom right panel). It is clear from the figure, especially when the binaries host only fully convective stars (top left panel), that the strength of magnetic braking does not strongly contribute to shape the distributions. On the other hand, the orbital period distribution of main-sequence binaries is strongly affected by how the magnetic braking torque depends on the star spins as shown by \citet{ElBadry_2022}.}
  \label{FigMSMS}
\end{figure*}

\begin{figure*}
\begin{center}
\includegraphics[width=0.999\linewidth]{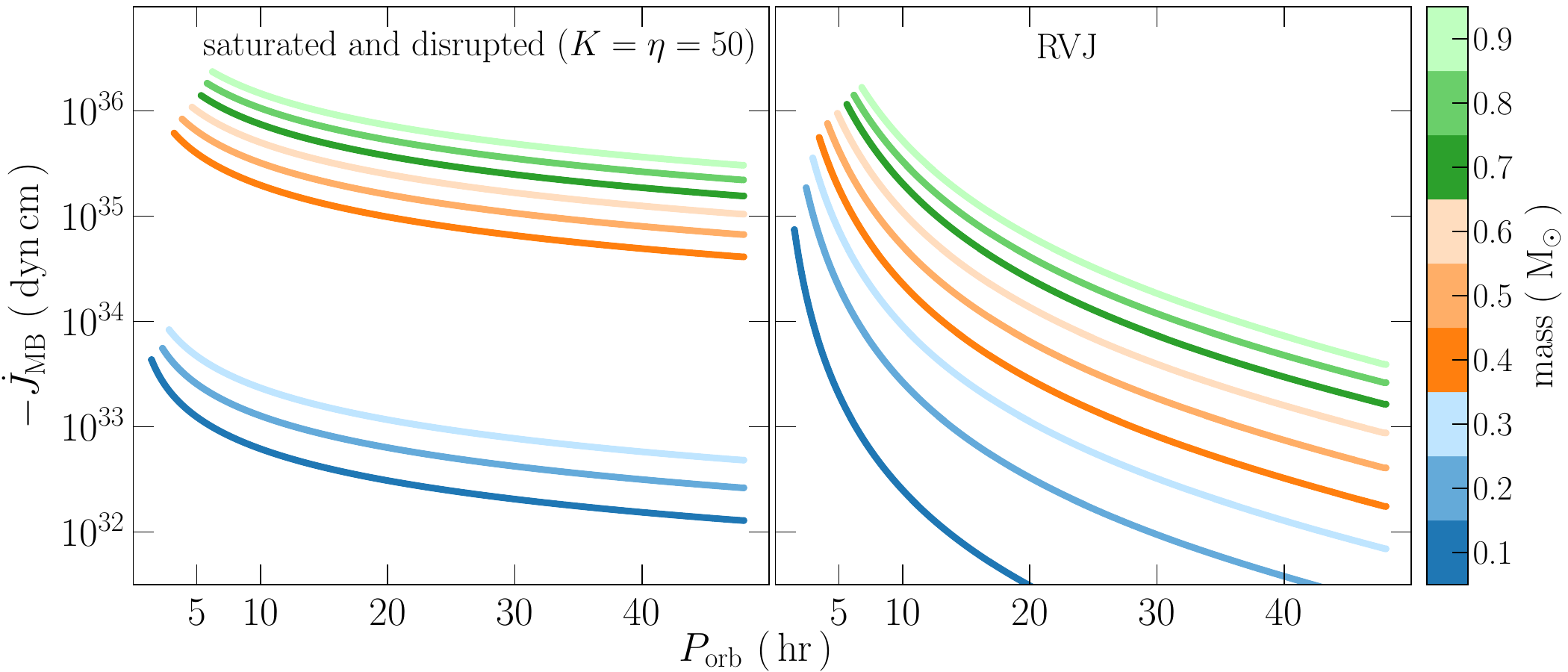}
\end{center}
\caption{Angular momentum loss rates obtained with the often adopted RVJ prescription (right panel) and the saturated and disrupted prescription (Eq.~\ref{MBrecipe}), assuming ${K=\eta=50}$ (left panel), for different main-sequence star masses and fixed white dwarf mass of $0.6$~\Msun. The torques provided by the RVJ prescription strongly depend on the orbital period, unlike the saturated prescription. For main-sequence stars with radiative cores that are near Roche lobe filling, both prescriptions lead to comparable angular momentum loss rates. For fully convective main-sequence stars, magnetic braking according to the saturated and disrupted prescription is order of magnitude weaker in comparison with the non-fully convective counterparts.}
  \label{FigMB}
\end{figure*}

We show in Figs.~\ref{FigPCEBflat} and \ref{FigPCEBmd17} the fractions of PCEBs amongst white dwarf plus M dwarf binaries as function of the M dwarf mass for several combinations of the model parameters $K$ and $\eta$ and for the two IBPs we adopted and compare with the observationally-inferred fractions \citep{schreiberetal10-1}.
The observational and all predicted distributions are normalized at their highest fractions since our goal is to reproduce the qualitative shape of the observed distribution as well as the relative changes in the fractions across the M dwarf mass.
From this comparison we can derive clear constraints on the assumed IBPs as well as the strengths and mass dependencies of magnetic braking.

%----------------------
% NEW SUBSECTION
%----------------------
\subsection{Importance of the dearth of extreme mass ratio binaries}

The observed fraction of PCEBs remains roughly constant for M dwarfs masses smaller than ${\sim0.3}$~\Msun, that is, for fully convective M dwarfs.
From this point on, the larger the M dwarf mass the lower the fraction of PCEBs.

Assuming uncorrelated flat IBP (Fig.~\ref{FigPCEBflat}), we can clearly see that the high fraction of PCEBs with M dwarf of masses ${\sim0.1-0.2}$~\Msun~cannot be reproduced, irrespective of the assumed strength of magnetic braking. 
While the observed fraction remains very high for the lowest mass main-sequence stars, the simulations predict a significant decrease of the fractions of PCEBs towards lower main-sequence star masses.

This decrease is easy to understand.
Assuming a flat IBP, for M dwarf masses of ${\sim0.1-0.2}$~\Msun, most white dwarf plus M dwarf binaries have progenitor mass ratios smaller than ${\sim0.1}$.
The initially very wide systems of these extreme mass ratio systems, that is, those in which both stars do not interact, become wide white dwarf plus main-sequence binaries regardless of the main-sequence star mass. 
In contrast, the fraction of PCEB progenitors with such extreme mass ratios that survive common-envelope decreases with decreasing mass of the main-sequence star.
This lowers the number of PCEBs towards smaller masses and in turn the fraction of PCEBs with respect to all white dwarf plus main-sequence binaries.
Given that this feature of the predicted distributions occurs independent on the assumed magnetic braking parameters, it indicates that the assumed flat IBP is probably wrong.

Indeed, when comparing the results of both IBPs, we can see that there is no problem at all to reproduce the high fraction of PCEBs with M dwarf companions of masses ${\sim0.1-0.2}$~\Msun~in the case of the \MDt~IBP (Fig.~\ref{FigPCEBmd17}).
This is a consequence of the previously mentioned lower limit in the mass ratio distribution at ${q=0.1}$ in this IBP, which is motivated by observations \citep[e.g.][]{Grether_2006,Moe_2015a,Murphy_2018}.
In other words, the PCEB fraction among white dwarf plus main-sequence binaries with low-mass M dwarfs can be explained if the dearth of extreme mass ratio systems, for solar type stars often called the brown dwarf desert, is taken into account.

One may think that the lack of binaries with ${q<0.1}$ in the IBP should reduce both the number of PCEBs with low-mass M dwarfs and the number of wide systems with low-mass M dwarfs, and for this reason there should be no differences at all between both IBPs adopted here.
However, by removing ${q<0.1}$ binaries from the IBP, those binaries that are progenitors of PCEBs with low-mass M dwarfs can more easily survive CE evolution since these low-mass M dwarfs are on average more massive.
This removal of ${q<0.1}$ binaries results in a reduction in the number of PCEBs with low-mass M dwarfs (a factor of ${\lesssim4}$) that is smaller than the reduction in the number of wide systems (a factor of ${\sim10}$).
This leads in turn to the above mentioned agreement between prediction and observations.
We confirmed that by running the same models for the flat IBP but enforcing that ${q>0.1}$ always, and we obtained fractions similar to those shown in Fig.~\ref{FigPCEBmd17}.

%----------------------
% NEW SUBSECTION
%----------------------
\subsection{Evidence for disrupted magnetic braking}

Towards main-sequence star masses larger than $0.2$~\Msun, the observed fraction of PCEBs continuously decreases.
At an M dwarf mass of ${\sim0.5}$~\Msun, the observed fraction of PCEBs has dropped by a factor of ${5-10}$ compared to systems with low-mass M dwarfs.

The observed fraction of PCEBs with M dwarfs of masses ${\gtrsim0.3}$~\Msun~can be fairly well reproduced regardless of the adopted IBP, as long as the magnetic braking torque is sufficiently strong, which occurs when ${K\gtrsim50}$, and drops at the fully convective boundary by at least a factor of ${\eta\sim50}$.  
This result holds for both IBPs because, in this mass range, the common-envelope survival rate is not dramatically affected by the number of binaries with extreme mass ratios.

Therefore, the drop of the fraction of PCEBs with respect to white dwarf plus main-sequence binaries is consistent with the disrupted magnetic braking scenario, i.e. because the orbital angular momentum loss due to magnetic braking drastically decreases at the fully convective boundary.
Efficient magnetic braking causes zero-age PCEBs with main-sequence stars that have a radiative core to evolve into CVs on a short time-scale compared to PCEBs with fully convective main-sequence stars. 
We will discuss in more detail in what follows how $K$ and $\eta$ shape the distribution of the fraction of PCEBs across the M dwarf mass using the outcomes when the \MDt~IBP is adopted.

%----------------------
% NEW SUBSECTION
%----------------------
\subsection{Evidence for strong and weak magnetic braking}

Comparison with the observations does not only provide evidence for disrupted magnetic braking for fully convective main-sequence stars but also an increased strength of magnetic braking for main-sequence stars with a radiative core.
Assuming ${K=1}$ or ${K=10}$ (top panels of Fig.~\ref{FigPCEBmd17}), the high fractions of PCEBs hosting fully convective M dwarfs can be nicely reproduced for all values of $\eta$, except ${\eta=1}$.
However, even no magnetic braking (${\eta\rightarrow\infty}$) does not provide evolutionary time scales different enough to reproduce the decrease at of the fraction of PCEBs at the fully convective boundary.

Only if the strength of magnetic braking is increased by a factor of at least ${K\sim50}$ and disrupted (i.e. ${\eta\gtrsim50}$), the relative numbers of PCEBs with fully convective M dwarfs and the number of those with more massive M dwarfs agree with the observations. 
In other words, magnetic saturation alone cannot account for the relatively larger number of PCEBs with fully convective M dwarfs in comparison with PCEBs with more massive M dwarfs.
Magnetic braking needs to be stronger (${K\gtrsim50}$) for main-sequence stars with radiative cores than provided by the standard saturated magnetic braking prescription (i.e. Eq.~\ref{MBsat}).

%----------------------
% NEW SUBSECTION
%----------------------
\subsection{Main-sequence binaries} 

In the previous subsections we have provided clear evidence for disrupted magnetic braking from observations of close binary stars. 
At this point one might be wondering whether the constraints for the detached eclipsing low-mass main-sequence binaries provided  by \citet{ElBadry_2022} would be violated or not by arbitrarily changing the strength of magnetic braking.

We carried out main-sequence binary population synthesis with the assumptions described in Sect.~\ref{BSE}.
We picked the primary mass from the canonical \citet{Kroupa_2001} initial mass function and the secondary from a uniform mass ratio distribution, assuming a minimum mass of $0.1$~\Msun.
The orbital period was also picked from a uniform distribution assuming a maximum of $5$~d and a minimum corresponding to a separation equal to $1.1$ times the sum of the primary and secondary radii.
The orbit was assumed to be circular, which is consistent with the strong tidal interaction expected to take place in such close binaries.
The age of each binary was chosen from a uniform distribution extending up to $10$~Gyr.
To better compare with \citet{ElBadry_2022}, we split the IBP into four sub-IBPs, each of which having $2\times10^5$ main-sequence binaries, according to the primary mass, namely
(i) ${0.10-0.30}$~\Msun, 
(ii) ${0.30-0.40}$~\Msun,
(iii) ${0.40-0.65}$~\Msun, and 
(iv) ${0.65-0.90}$~\Msun.

We compare in Fig.~\ref{FigMSMS} predicted and observed distributions fixing ${K=50}$.
Our simulations fit the observations as well as the one based on saturated magnetic braking by \citet{ElBadry_2022}.
This is not surprising because the normalized period distribution of main-sequence binaries only constrains the dependence of magnetic braking on the spin period which is identical in our prescription and any other saturated magnetic braking prescription.

%%%%%%%%%%%%%%%%%%%%%%%%%%%%%%%%
%%%%%%%%%%%%%%%%%%%%%%%%%%%%%%%%
% NEW SECTION
%%%%%%%%%%%%%%%%%%%%%%%%%%%%%%%%
%%%%%%%%%%%%%%%%%%%%%%%%%%%%%%%%
\section{Discussion}
\label{Discussion}

We have combined two very clean constraints on orbital angular momentum loss through magnetic braking (observations of detached eclipsing main-sequence binaries and detached white dwarf plus M dwarf binaries) and found a purely empirical prescription for magnetic braking that can reproduce both observations.
To explain the observed PCEB distribution, the strength of magnetic braking needs to significantly change at the fully convective boundary.
We need a ${\gtrsim50}$ times stronger magnetic braking for stars that still contain a radiative core compared to fully convective stars to explain the dramatically increased fraction of PCEBs among white dwarf plus M dwarf binaries with fully convective main-sequence stars \citep[see also][]{schreiberetal10-1}.

To also reproduce the flat period distributions observed of main-sequence binaries for all type of main-sequence M dwarf binary stars \citep{ElBadry_2022}, both magnetic braking prescriptions need to weakly depend on the orbital period as predicted by the saturated magnetic braking prescriptions \citep[e.g.][]{Sills_2003}.
Combining both these constraints leads to a prescription that can simultaneously explain both observational facts.
In what follows we briefly discuss to which degree alternative explanations of the observations might exist and the implications of our results for other types of close binary stars.

%----------------------
% NEW SUBSECTION
%----------------------
\subsection{Do reasonable alternative explanations exist?}

If the significant decrease of the fraction of PCEBs among white dwarf plus M dwarf binaries at the fully convective boundary was not caused by a dramatic change in the efficiency of orbital angular momentum loss through magnetic braking, the observations would need to be explained by previous evolutionary effects that make the formation of PCEBs with M dwarfs with a radiative core unlikely.
This would imply that for some reason, common-envelope evolution leads to the merger or very short post-common-envelope orbital periods of white dwarf plus early M dwarf companions while fully convective stars are more likely to emerge at longer periods.

This possibility,  however, appears to be very unlikely.
First, for a given primary mass and orbital period, the available orbital energy is larger for more massive secondary stars which makes it actually easier to survive the common-envelope phase.
Second, we do not see evidence for a relation between common-envelope efficiency and main-sequence star mass in the observed samples of PCEBs \citep{Zorotovic_2014,Zorotovic_2022}, and last but not least we find a significant number of descendants from PCEBs with early low-mass main-sequence stars in observed samples of CVs \citep{Pala_2020,Pala_2022}.

Concerning the main-sequence binary orbital period distribution as measured by \citet{ElBadry_2022}, we do not see any reasonable alternative explanation than a magnetic braking prescription that depends weakly on the spin period.
One could in principle think of a flat birth distribution combined with extremely weak magnetic braking but this would not only disagree strongly with the fraction of PCEBs but also with observations of the spin down rates of single stars \citep[e.g.][]{newtonetal16-1}.
It therefore appears to us that the two samples we analysed in this work provide solid evidence for a disrupted and saturated magnetic braking prescription for binaries hosting main-sequence stars with masses between $\sim0.1-0.9$~\Msun.

%----------------------
% NEW SUBSECTION
%----------------------
\subsection{Implications for single stars and other binary stars}

The prescription of a disrupted and saturated magnetic braking we derived is purely empirical and based on observations of binary stars with main-sequence stars less massive than $\sim0.5$~\Msun~and $\sim0.9$~\Msun~in the case of the PCEBs and main-sequence binary stars samples, respectively. 
In CVs, a white dwarf accretes hydrogen rich material from a low-mass main-sequence star and our empirical prescription should therefore properly describe angular momentum loss through magnetic braking in CVs.
We think that this is most likely the case.

The decrease of at least a factor of ${\sim50}$ at the fully convective boundary will clearly generate a period gap, that is, a  dearth of semi-detached systems with orbital periods between two and three hours.
This gap occurs because of the following.
Non-fully convective main-sequence stars are driven out of thermal equilibrium due to strong magnetic braking and the correspondingly large mass transfer rate and evolve towards shorter orbital periods.
As soon as the main-sequence star becomes fully convective, at an orbital period of about three hours, magnetic braking becomes much less efficient, the main-sequence star has enough time to relax to its equilibrium radius and consequently does not continue to fill its Roche lobe, and the binary evolves as a detached binary until mass transfer resumes at an orbital period of about two hours.
The increased main-sequence radius above the gap 
\citep[][]{Knigge_2011_OK} as well as detached CVs crossing the gap have been observed \citep{Zorotovic_2016}.
The period gap itself is evident in large magnitude limited samples \citep[][]{Knigge_2011_OK} but hardly visible in volume limited samples \citep{Pala_2020}, which is likely caused by the small sample size.

Our model should also be able to reasonably well produce mass transfer rates similar to the observed ones above and below the period gap as magnetic braking for main-sequence stars with radiative cores is much stronger than that for fully convective main-sequence stars (Fig.~\ref{FigMB}).
However, in general, we advocate caution when using mass transfer rates derived from observations of CVs to constrain magnetic braking prescriptions.
This is because secular mass transfer rates are rather difficult to measure.
The radii of the main-sequence stars \citep{Knigge_2011_OK,McAllister_2019} and the white dwarf effective temperatures \citep{palaetal_2017,Pala_2022} have been used but for systems with non-fully convective main-sequence stars only a handful of reliable measurements exist.
We finally note that weak and continuous saturated magnetic braking as suggested previously will neither produce mass transfer rates similar to the observed ones nor lead to the prediction of an orbital period gap unless both features are produced mainly by consequential angular momentum loss. 
This though seems very unlikely as consequential angular momentum loss is regulated by the mass transfer process itself and for this reason cannot drive CV evolution.

Other types of close binaries whose evolution is largely influenced by angular momentum loss through magnetic braking are progenitors of AM\,CVn binaries, low-mass X-ray binaries, ultra-compact X-ray binaries, and progenitors of close binaries hosting millisecond pulsars paired with helium white dwarfs.
The characteristics of the observed samples of these binaries seem to be explained best assuming a convection and rotation boosted (CARB) magnetic braking \citep{vanetal19-1,vanetal19-2,chenetal21-1,Soethe2021,Deng2021,BelloniSchreiber2023} for stars that are nuclear evolved and more massive than ${\sim1}$~\Msun.
While our prescription predicts relatively strong magnetic braking for stars with a radiative core and a convective envelope, one would need to run dedicated simulations to find out whether it can explain the observations as well as results achieved with the CARB prescription.
Alternatively, our prescription could only be valid for lower mass stars (${\lesssim0.9}$~\Msun) while the CARB model applies for nuclear evolved and more massive stars ($\gtrsim1$~\Msun).
In case of the latter, one would need to understand the physical reasons behind this apparent switch in the strength and dependencies of magnetic braking.

The situation is equally complicated for single stars.
Braking prescriptions derived from rotation rates of single M dwarfs in star clusters have been claimed to not show a discontinuity at the fully convective boundary \citep[see][their section 6.2]{godoy-riveraetal21-1} which, if true, contradicts our finding.
Taking a detailed look at the K2 data for the Praesepe cluster, \citet{douglasetal17-1} found indications for an increased braking efficiency for early M dwarfs relative to solar-type stars and fully convective stars which agrees with our prescription. 
Samples of older field M stars indicate that fast rotators are far more frequently found among fully convective stars \citep{newtonetal16-1} which could agree with our prediction of less efficient braking in these stars.
Recently, \citet{passetal22-1} analysed the rotation of M dwarfs that are members of wide binary stars which allowed an age determination and found that fully convective stars remain fast rotators up to an age of ${2-3}$~Gyr but spin down at later stages.
This has been interpreted as a time dependence in the strength of angular momentum loss in young single stars \citep{brown14-1}.
However, old M dwarfs experiencing spin–orbit interactions as members of unresolved binaries are forced to rapid rotation and activity persist \citep{passetal22-1}.
This might indicates that the results from the spin-down of single stars might not be applicable to close binary stars.

We conclude that the empirical braking laws we derive from two samples of detached binary stars are a suitable magnetic braking prescription for close binaries with stars less massive than ${\lesssim0.9}$~\Msun.
To what degree the same braking can be applied to more massive potentially nuclear evolved stars in binaries or to single stars of the same mass remains highly uncertain.

%%%%%%%%%%%%%%%%%%%%%%%%%%%%%%%%
%%%%%%%%%%%%%%%%%%%%%%%%%%%%%%%%
% NEW SECTION
%%%%%%%%%%%%%%%%%%%%%%%%%%%%%%%%
%%%%%%%%%%%%%%%%%%%%%%%%%%%%%%%%
\section{Conclusions}\label{Conclusion}

We have performed population synthesis for white dwarf plus M dwarf binary stars and found that:

\begin{description}
    \item[i)] a magnetic braking prescription that is disrupted, i.e drops by a factor of ${\gtrsim50}$ at the fully convective boundary, can explain the observed drop of the fraction of post-common-envelope binaries at the fully convective boundary; 
    \item[ii)] magnetic braking needs to be stronger than assumed in standard prescriptions of saturated magnetic braking for M dwarfs with a radiative core and weaker in the case of fully convective main-sequence stars;
    \item[iii)] the observed large fraction of post-common-envelope binaries with low-mass main-sequence stars (${\sim0.1-0.2}$~\Msun) is naturally reproduced if the dearth of extreme mass ratio binaries among main-sequence binaries is taken into account in the initial binary distributions, that is,  the brown dwarf desert observed at separations ${<1}$~au must extend to wider separations ${\sim1-10}$~au, albeit not as dry as observed at very close separations. 
\end{description}

As we assumed that in addition to being disrupted, magnetic braking is saturated, i.e depends weakly on the orbital period, the predictions of our prescription is in reasonable agreement with the period distribution of main-sequence binary stars.
We discussed possible alternative explanations for the observations but did not find a convincing one.
We therefore conclude that saturated and disrupted magnetic braking represents an adequate magnetic braking law for ${\sim0.1-0.9}$~\Msun~main-sequence stars that are members of close binary systems.
However, a physically rather than empirically motivated saturated magnetic braking law is required to eventually understand magnetic braking.

%%%%%%%%%%%%%%%%%%%%%%%%%%%%%%%%
%%%%%%%%%%%%%%%%%%%%%%%%%%%%%%%%
% NEW SECTION
%%%%%%%%%%%%%%%%%%%%%%%%%%%%%%%%
%%%%%%%%%%%%%%%%%%%%%%%%%%%%%%%%
\begin{acknowledgements}

We would like to thank an anonymous referee for the comments and suggestions that helped to improve this manuscript.
We thank the Kavli Institute for Theoretical Physics (KITP) for hosting the program ``White Dwarfs as Probes of the Evolution of Planets, Stars, the Milky Way and the Expanding Universe'' and the Munich Institute for Astro-, Particle and BioPhysics (MIAPbP) for hosting the program ``Stellar Magnetic Fields from Protostars to Supernovae''.
This research was supported in part by the National Science Foundation under Grant No. NSF PHY-1748958 and by the Munich Institute for Astro-, Particle and BioPhysics (MIAPbP) which is funded by the Deutsche Forschungsgemeinschaft (DFG, German Research Foundation) under Germany's Excellence Strategy – EXC-2094 – 390783311.
We thank Jim~Fuller for pleasant and helpful discussions during the KITP program.
DB acknowledges financial support from {FONDECYT} grant number {3220167}.
MRS was supported from {FONDECYT} grant number {1221059} and ANID, – Millennium Science Initiative Program – NCN19\_171.
KJS was supported by NASA through the Astrophysics Theory Program (80NSSC20K0544).

\end{acknowledgements}


\bibliographystyle{aa} % style aa.bst
\bibliography{references} % your references Yourfile.bib

\end{document}